# A Strained Organic Field-Effect-Transistor with a Gate-Tunable Superconducting Channel

Hiroshi M. Yamamoto[1,2,3], Masaki Nakano[4,5], Masayuki Suda[1,2], Yoshihiro Iwasa[5,6], Masashi Kawasaki[5,6], and Reizo Kato[2].

[1] Research Center of Integrative Molecular Systems (CIMoS), Institute for Molecular Science, 38 Nishigounaka, Myodaiji, Okazaki, Aichi 444-8585, Japan.

[2] RIKEN, 2-1 Hirosawa, Wako, Saitama 351-0198 Japan

[3] JST, PRESTO, 4-1-8 Honcho, Kawaguchi, Saitama, 332-0012 Japan (PRESTO: Precursory Research for Embryonic Science and Technology)

[4] Institute for Materials Research, Tohoku University, 2-1-1 Katahira, Aoba-ku, Sendai 980-8577, Japan

[5] RIKEN Center for Emergent Matter Science, 2-1 Hirosawa, Wako, Saitama 351-0198 Japan

[6] Quantum Phase Electronics Center and Department of Applied Physics, School of Engineering, The University of Tokyo, 7-3-1 Hongo, Bunkyo-ku, Tokyo 113-8656, Japan

Correspondence and requests for materials should be addressed to H.M.Y. (email: yhiroshi@ims.ac.jp)

In state-of-the-art silicon devices, mobility of the carrier is enhanced by the lattice strain from the back substrate. Such an extra control of device performance is significant in realizing high performance computing and should be valid for electric-field-induced superconducting devices, too. However, so far, the carrier density is the sole parameter for field-induced superconducting interfaces. Here we show an active organic superconducting field-effect-transistor whose lattice is modulated by the strain from the substrate. The soft organic lattice allows tuning of the strain by a choice of the back substrate to make an induced superconducting state accessible at low temperature with a paraelectric solid gate. An active three terminal Josephson junction device thus realized is useful both in advanced computing and in elucidating a direct connection between filling-controlled and bandwidth-controlled superconducting phases in correlated materials.

## Introduction

Strained silicon shows enhanced mobility due to the band-structure tuning and is widely used in modern high-spec circuits[1]. This technique should be also applicable to the innovation of wider range of electronic devices including a field-induced superconducting (SC) interface[2-6]. If a strained SC field-effect-transistor (FET) is realized, it will contribute in finding new materials for, in figuring out unknown phase diagrams of and in utilizing quantum devices based on, superconductivities. In general, organic materials enjoy soft lattice and therefore the strain effect on its properties should be significant because of the alteration in the non-covalent intermolecular interaction. Indeed, the organic superconductivity is known to be sensitive to the physical and chemical pressure effects, providing a well-investigated platform to check the influence of strain on the SC transition. An organic superconductor that neighbors Mott-insulating (MI) phase has several advantages in realizing SC-FET with a strained interface with following reasons. Firstly, the relationship between the strain and the electron correlation $U/W$, where $U$ is the effective Coulomb interaction and $W$ is the bandwidth, is well established[7,8], leading to precise analysis of the obtained result. Secondly, the relationship between $U/W$ and SC state in the phase diagram with half-filled (non-doped) band is already well-investigated both experimentally and theoretically[9-12]. Thirdly, the carrier number required for the induction of SC state is expected to be relatively low because of the large molecular size along with the dimerized crystal structure. For example, the half-filled carrier density for the κ-type BEDT-TTF (BEDT-TTF = bis(ethyelenedithio)tetrathiafulvalene) material (*ca.* $2 \times 10^{14}$ / cm$^2$)[13] is less than one third of that for high-Tc cuprates (*ca.* $7\times10^{14}$ / cm$^2$), which will make a field-induced superconductivity within the reach of carrier density modulation with solid paraelectric gate insulator, and thus an active control of the device in a real-time low temperature experiment is possible. And finally, there should be no dangling bond and associated carrier trap on the surface because molecular crystal is constructed only by weak intermolecular interactions, which is good for transistor operation.

The above four reasons justify a strained organic SC-FET as an ideal test ground for the strain effect in gate-controlled SC devices. At the same time, SC-FET with organic material merits, when it is realized, also in understanding the phase diagram of correlated materials' superconductivity in simultaneous control of bandwidth and bandfilling, which has been impossible to be studied by other methods.  For example, bandwidth-controlled SC phase with half-filled condition has not yet been realized in cuprates because of its hard lattice, although it is important in verifying the superconductivity mechanism to know whether filling-controlled and bandwidth-controlled SC phases are connected to one another or not.

In this paper, we demonstrate an active field-effect control of superconductivity in a strained FET with an organic Mott insulator, κ-(BEDT-TTF)$_2$Cu[N(CN)$_2$]Br (κ-Br), whose ground state is tuned in the vicinity of a strain-induced Mott-transition. This device provides a novel three-terminal Josephson junction whose transport characteristics reveals a phase diagram of an organic Mott-insulator, where a direct connection between filling-controlled and bandwidth-controlled SC phases is suggested.

## Results

### Strain effect from the substrate

We fabricated FETs by laminating thin single crystals of κ-Br on top of metallic Nb-doped $SrTiO_3$ (STO) substrates covered with 30 nm of $Al_2O_3$ dielectric layer grown by atomic layer deposition (Fig. 1a). κ-Br is a highly correlated organic superconductor ($T_c$ = 11.6 K)[14] whose bandwidth-controlled SC phase neighbors an antiferromagnetic MI phase[11] (Fig. 1b). The ground state of κ-Br can be finely tuned by physical and/or chemical pressure. In our previous studies[15], by laminating it onto $SiO_2$/Si substrates, we applied tensile strain to thin crystals of κ-Br and guided them into a MI state at low temperature. Because thermal expansion coefficients are different between a Si substrate and a κ-Br crystal (ca. 2 and 60 ppm / K at room temperature, respectively), the κ-Br crystal is 'expanded' at low temperature as shown in Fig. 1b, blue arrow. In the present study, STO was chosen as a substrate because of its relatively large thermal expansion coefficient (about 10 ppm / K at room temperature) in order to adjust the ground state of κ-Br very close to a partially SC region by a weaker tensile strain. In Fig. 1c, we show normalized temperature dependence of the resistance ($R$-$T$ plot) for two devices [**1** (orange) and **2** (red)] as well as those for κ-Br bulk (black) and κ-Br on a $SiO_2$/Si substrate (blue). Because of the weak tensile strain effect from STO substrate, the resistances for these two samples remain in between a complete SC state and a highly-resistive MI state, as indicated in Fig. 1b by orange and red arrows. Device **1** showed insulating behavior at low temperature, while device **2** showed a small resistance drop at 12 K, followed by a reentrant percolation transition around 9 K. This behavior of device **2** indicates that the system is in the partially SC phase where separation between SC and MI phases occurs and the SC fraction is maximized around 9 K. Such a situation has been already investigated in detail for a bulk material by infrared spectroscopy mapping[16], nuclear magnetic resonance[17], noise- measurement[18] and so on. In our devices, magnetization measurement for another sample (device **3**) also showed partial superconductivity and maximization of SC fraction at medium temperature range, which will be described later. The above sample dependence between devices **1** and **2** indicates that the conductivity of κ-Br is very sensitive to the small difference in the strain that is produced in the lamination and cooling processes.

### N-type field-effect and superconductivity in device 1

By applying gate voltage ($V_G$), the resistance of the device **1** drastically changed with '*n*-type' polarity as shown in Fig. 2a. With negative $V_G$, the resistance increases rapidly, while application of positive $V_G$ squeezed out the insulating phase into low temperature region and the device became weakly metallic at $V_G = 2$ V. By further increasing $V_G$, it showed superconductivity at $V_G > 8$ V. The current-voltage characteristic in the SC region (Fig. 2a inset) showed typical bistable switching of a resistance-shunted Josephson junction (JJ) with a McCumber parameter much larger than unity[19], which implied an inhomogeneous SC transition at the interface. The contour map of the resistance in logarithmic scale is shown in Fig. 2b, suggesting a SC dome very similar to that for cuprate superconductors (white dashed line). This evidences that the cuprates and the κ-type organic system share the common phase diagram in the filling-controlled regime.

### P-type field-effect and superconductivity in device 2

In Fig. 3a we show an $R$-$T$ plot below 13 K for device **2**. The device entered partially SC state at 12 K even at $V_G$ = 0 V, followed by a reentrant percolation transition around 9 K (OFF → ON transition at 9.2 K, and ON → OFF transition at 8.1 K). This reentrant transition is already known in a κ-type bulk crystal in the vicinity of bandwidth-controlled Mott-transition[21], and can be explained by the recurrent decrease of SC volume fraction at low temperature. A schematic image of a percolation transition of a JJ network (JJN) in our device is shown in Fig. 3b. The 'ON' switching of JJN designates the minimum of the free energy for the SC state with respect to that for the MI state, because the fraction ratio of these two competing phases reflects the relative difference in their free energies. The JJN in device **2** was switchable *in-situ* by changing either magnetic or electric fields, too: When the magnetic field was swept, the device exhibited switching with large hysteresis (ON → OFF at 1.2 T, OFF → ON at 0.8 T; Fig. 3a inset). As for the gate electric field, the device exhibited reentrant JJN switching with '*p*-type' polarity upon $V_G$ sweeping below 8 K (Fig. 3c). The $V_G$ required for the switching was shifted to larger negative value as the temperature decreased. The whole resistance mapping as a function of $T$ and $V_G$ is shown in Fig. 3d. In a mid-gap $V_G$ area (−4 V <

$V_G$ < 11 V), where essentially no mobile carriers were introduced due to mid-gap traps (*i.e.* the essential bandfilling is just 0.5), the JJN was switched to the ON state around 8 - 9 K (blue area). At $V_G$ < −4 V, on the other hand, the switching temperature started shifting downward as the hole injection proceeded. The blue area continued to reach to a fully hole-doped SC state around ($T$, $V_G$) = (3 K, −13 V). Because these JJN switchings at around ($T$, $V_G$) = (8.5 K, 0 V) and ($T$, $V_G$) = (3 K, −13 V) reflect the free energy minima for the bandwidth-controlled and the filling-controlled SC states, respectively, Fig. 3d clearly shows a seamless connection between these two different SC states.

### Magnetic measurement for n-type device 3

We also confirmed by magnetic susceptibility measurement (for device **3**, Fig. 4) that the devices exhibited diamagnetic shielding effect whose volume fraction could be modulated by applying $V_G$. Although there was strong sample dependence, device **3** exhibited the largest Meissner and shielding effect among several samples we have measured, with the magnetic field applied parallel to the BEDT-TTF layers. At $V_G$ = 0 V, the OFF-state shielding effect was observed below 7 K, while the shielding effect was enhanced by applying a positive $V_G$, implying the device switching to the

ON state with the $n$-type characteristics. This enhancement was reproducible at $V_G$ sweep. With an assumption that the shielding effect is proportional to the volume fraction of the superconductivity and that the volume fraction of the (unstrained) bulk crystal is 100 %, one can estimate the SC fraction of both OFF and ON states of the device, by comparing the diamagnetic values of device **3** with that of bulk κ-Br. As a result, the temperature dependency of the volume fraction of superconductivity in device **3** at various $V_G$ is estimated as shown in Fig. 4b. Taking into account that the thickness of the charge accumulation layer is one- or two-layers[15] and the thickness of κ-Br crystal on the device **3** (550 nm) corresponds that of about 350 BEDT-TTF layers, the SC transition seems not only to take place in the charge-injected layers but also to propagate into the neighboring layers in the thickness direction, because *ca.* 5 % the change of the volume fraction between OFF ($V_G$ = 0 V) and ON ($V_G$ = 10 V) states well exceeds the expected value for monolayer transition (*ca.* 0.3 %). Such a bulk phase transition seems to originate from an interlayer dielectric screening of the Coulomb interaction $U$[20]. The maximization of the SC fraction (or, recurrent decrease of SC fraction) around 5 K of device **3** also designates a characteristic feature of κ-Br system in this percolative region, which is commonly seen in the reentrant percolation transition of device **2**.

## Discussion

So far, the hole- or electron-doped cuprates' superconductivity and κ-type organic superconductivity have been separately discussed as in Fig 5a, although scientists have pointed out many common physical properties[22]. These similarities are a consequence of a Mott-transition, where doping controls the bandfilling of cuprates while a pressure/strain controls bandwidth of organic Mott-insulators. However, the simultaneous control of the strain (bandwidth) and the carrier density (bandfilling) in a single Mott device is necessary for a deeper discussion of SC-MI phase competition in the unified diagram. For example, the order parameter symmetry is pointed out to be $d_{x^2-y^2}$ for cuprate and $d_{xy}$ for κ-BEDT-TTF.[23, 24] In order to discuss the origin of this discrepancy, it is important to realize both filling-controlled and bandwidth-controlled superconductivity in the same material and map out the connection between these two phases, which has been impossible to be checked with cuprates that exhibits a hard lattice. In the present experiments, it turned out that these SC phases are directly connected to one another as drawn in Fig. 5b. Inside the yellow SC arc, the $U/W$ is too high to evoke superconductivity, while it is too weak outside the arc. The present organic FET devices having controllable bandwidth and bandfilling therefore provide an indispensable opportunity for exploring a phase diagram of a channel material in a wide parameter space.

Judging from the $R$-$T$ plots for devices **1** and **2**, the difference in the strain for these devices corresponds to *ca.* 10 MPa difference in hydrostatic pressure for bulk material.[21] From crystallographic experiment, the lattice of κ-type BEDT-TTF material is known to shrink at a rate of about 0.025 % per 10 MPa, and the transfer integrals between molecules increases at a rate of about 0.2 % per 10 MPa.[25] This means that the difference in $U/W$ for devices **1** and **2** (in Fig. 5b) is only 0.2 % with respect to the absolute value of $U/W$ (an order of unity), when one takes into account the fact that $W$ is proportional to the transfer integral ($W = 4|t|$ where $t$ is the inter-dimer transfer integral) while effective $U$ is not sensitive to the change in the transfer integral[7] (Because $U \approx 2|t_{dimer}| - 4t_{dimer}^2/U_{bare}$, where $t_{dimer}$ is intra-dimer transfer integral whose value is about 0.25 eV and $U_{bare}$ is on-site Coulomb repulsion for a single BEDT-TTF molecule whose value is about 1.0 eV, small changes in $t_{dimer}$ cancels out.). Since such a small strain of 0.025 % in the lattice evokes a drastic change in the conductivity behavior and associated FET characteristics, the present results have proved that the control of the strain is a significant tool for the quest for organic SC-FETs.

Because the gate insulator in the present device is made of a paraelectric solid, an active switching of JJ has become possible. Such an electrostatically switchable three-terminal JJ device may find applications in SC computation methods such as rapid single-flux-quantum[26] and quantum computers,[27] because these 'beyond CMOS' methods utilizes superconducting quantum interference device (SQUID). In addition, an *in situ* gate-sweep measurement at low temperature allows a detection of first-order (hysteric) transitions in the bandfilling-controlled regime. One example of such a transition is already visible in Fig. 3c, where the resistance showed a bistable behavior depending on the sweep direction just inside the SC region. This first-order phase transition is presumably relevant to the discrepancy in one-particle excitation spectrum of a Mott-insulator that is dependent on the sweep-direction of the correlation strength as anticipated by theories[28-30].

In summary, we have realized a strained organic FET with gate-tunable SC channel. The phase diagram obtained by the device operation shows a direct connection between bandwidth-controlled and filling-controlled SC phases around a MI phase, which has been impossible to realize in inorganic cuprates. The gate-tunable Josephson junction may be useful in 'beyond CMOS' quantum devices and detecting a hidden first-order phase transition in a filling-controlled regime.

## Methods

### Substrate preparation

A 30 nm-thick amorphous $Al_2O_3$ gate insulator was grown at 150 °C on a 0.05 wt% Nb-doped $SrTiO_3$ (001) single crystal substrate (Shinkosha Co., Ltd.), which was used as a bottom gate. A thickness of $Al_2O_3$ was evaluated by low-angle X-ray reflection measurement. A surface roughness was less than 0.2 nm, which is comparable to that of a substrate. Typical electrical properties of a $Al_2O_3$ dielectric layer were characterized by current-voltage and capacitance-voltage characteristics of a capacitor made of Au/Ti/$Al_2O_3$/Nb:$SrTiO_3$, showing large breakdown field (~ 6-7 MV/cm) and high dielectric constant (~ 9-10).

### Crystal growth and lamination process

All the chemical compounds were purchased from commercial sources and used without further purification unless noted. 1,1,2-trichloroethane was purified by a basic aluminum column. TTP[N(CN)$_2$] (TTP = tetraphenylphosphonium) was precipitated from a equimolar mixture of Na[N(CN)$_2$] and TTP-Br aqueous solutions, and recrystallized from ethanol-ethyl acetate. A thin (100-300 nm) single crystal of κ-Br was grown electrochemically by oxidizing BEDT-TTF (50 mg) dissolved in 100 ml of 1,1,2-trichloroethane (10 % v/v ethanol), in the presence of TTP[N(CN)$_2$] (200 mg), CuBr (50 mg), and TPP-Br (20 mg). After applying galvanostatic current of 5.0 μA for 15 hrs, thin crystal was picked up under microscopic observation. Then, it was transferred into an ethanol (10 ml) by pipette. A 3 mm square-STO substrate was immersed in the same ethanol and the crystal is guided on top of the substrate using the tip of a strand of hair. The substrate is then removed from the alcohol to be dried.

Transport measurements

The κ-Br crystal was cut by laser (V-technology; VL-C30-GB ) to form terminals. Gold wires (15 μmϕ) were attached to these terminals with silver and/or carbon paste. Standard four-probe measurement was done under He atmosphere in an automatic cryo-chamber (Niki-glass) equipped with a superconducting magnet.

Magnetic measurements

The magnetization of the device was measured by SQUID apparatus (Quantum

Design MPMS) equipped with reciprocating sample option. The weight of the thin κ-Br crystal on device **3** was calculated by using its thickness, area and specific gravity. Two phosphorous bronze wires were attached on the κ-Br and the gate substrate in order to modulate the gate voltage during the magnetic measurement. For field-cooled measurement, the device was cooled down from 20 K to 2 K, during which time a magnetic field of 8 T was applied, before the measurement. The magnetization was measured at a magnetic field of 100 Oe that is parallel to the BEDT-TTF layers in all the measurements.

**Acknowledgements**

We thank Y. Tokura, I. Kawayama and A. Fujimaki for discussions. This work was partly supported by Grant-in-Aid for Scientific Research (No.22224006, 21224009, 25708040, and 25000003) and the FIRST Program from JSPS and SICORP from JST.

**Author Contributions**

H.M.Y., M.N. and Y.I. designed the project. M.N. provided the substrates to H.M.Y. who grew the crystal, made the devices and measured the transport properties. M.S.

measured the magnetic susceptibility. All authors discussed the results and commented on the manuscript.

## Competing financial interests

The authors declare no competing financial interests.

**Figure legends**

Figure 1. Device configuration and the basic conduction properties.

**a,** Microscope image (upper right; scale bar, 100μm) and schematic section (lower right) of the device, along with crystal structure of κ-(BEDT-TTF)$_2$Cu[N(CN)$_2$]Br, where cationic BEDT-TTF layers and anionic Cu[N(CN)$_2$]Br layers alternate each other (left).

**b,** Phase diagram of κ-type BEDT-TTF system with bandfilling = 0.5, after ref. 17, 21, 31 and 32. The horizontal axis represents electron correlation which can be tuned by physical or chemical pressures as well as strain from the substrates. These pressure effects control bandwidth of the half-filled conduction band. PI, PM, AFI, and RN denote paramagnetic insulator, paramagnetic metal, antiferromagnetic insulator, and recurrent non-metallic state, respectively. Between AFI and complete SC, there is a percolative SC region where MI and SC phases coexist. Blue, orange, red, and black arrows indicate trajectories that κ-Br crystals on SiO$_2$/Si, on Al$_2$O$_3$/SrTiO$_3$(**1** and **2**), and as bulk experience upon cooling, respectively.

**c,** Temperature dependencies of relative resistance for κ-Br on SiO$_2$/Si (blue) or Al$_2$O$_3$/SrTiO$_3$ substrates (**1** orange, **2** red), as well as that of bulk crystal (black).

Figure 2. Characteristics for device **1**.

a, Temperature dependencies of the resistance of the device **1** under various $V_G$ in logarithmic scale. The inset shows $IV$ characteristics at $V_G = 9$ V and $T = 4$ K. The results can be repeated more than three times without any significant difference.

b, Contour plot of resistance as a function of temperature and $V_G$. The white broken line indicates the n-doped SC area. The upper horizontal scale shows the field-induced change in the electron number per BEDT-TTF dimer at the first monolayer interface.

Figure 3. Characteristics and switching mechanism for device **2**.

a, Temperature dependencies of device **2** resistance under various magnetic fields without $V_G$. A partial SC transition occurs at 12 K, after which percolation OFF/ON and ON/OFF transitions of JJN is observed. The inset shows resistance change under magnetic field sweep at $T = 8.5$ K, where percolation transition with large hysteresis is observed.

b, Schematic images representing OFF and ON states of the JJN, along with free energy diagrams for these two states. When the temperature, magnetic field, or $V_G$ is changed, the ratio of the SC fraction changes to evoke the JJN switching.

**c,** Resistance change under $V_G$ sweep at 13 K and 5 K. The arrows indicate the sweep directions.

**d,** Contour plot of resistance as a function of temperature and $V_G$ with rising $V_G$ sweep direction. These results (**a-d**) can be repeated more than three times without any significant difference.

Figure 4. Magnetic measurements for device **3**.

**a,** Magnetic susceptibility of device **3** under various $V_G$ (red, yellow, black, green and blue) along with that for bulk κ-Br (grey). FC and ZFC designate the measurements under field-cooled and zero-field-cooled conditions, respectively. The inset shows a magnified plot at low temperature region. Note that the inter-layer coupling in a bulk κ-Br crystal is a naturally formed Josephson coupling that allows penetration of magnetic fluxes to some extent.

**b.** Relative shielding fraction of device **3** in OFF ($V_G$ = 0 V) and ON ($V_G$ = 10 V) states calculated by the data in panel **a**. The shielding diamagnetism for device **3** is divided by that for the bulk κ-Br at the same temperature to estimate the volume fraction.

Figure 5. Phase diagrams around a Mott insulator.

**a,** Phase diagram that has been experimentally determined before this experiment.

**b,** Phase diagram that has been confirmed by this experiment. "SC (intrinsic)" denotes bandwidth-controlled SC phase, while "SC (n-dope)" and "SC (p-dope)" denote bandfilling-controlled SC phases. Devices **1** and **2** can be mapped as red arrows.

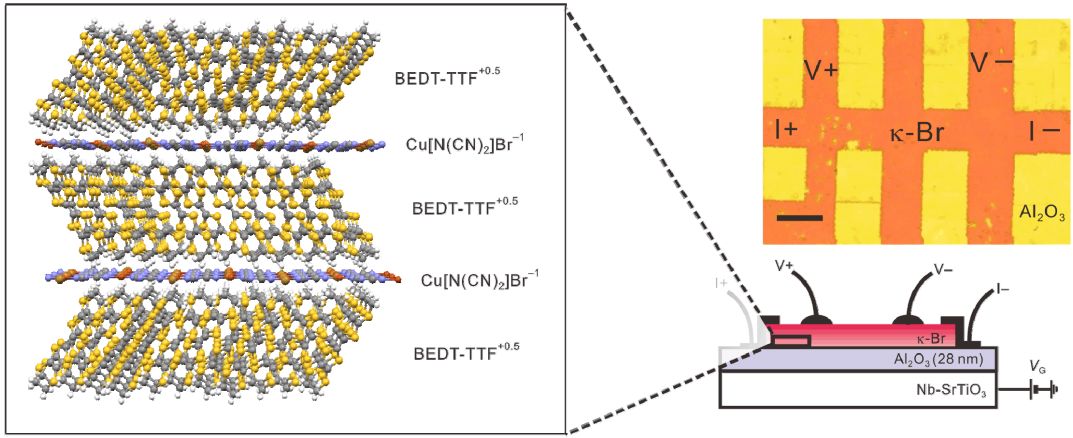

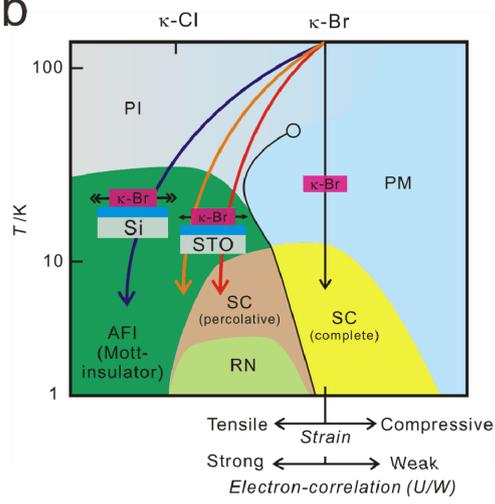
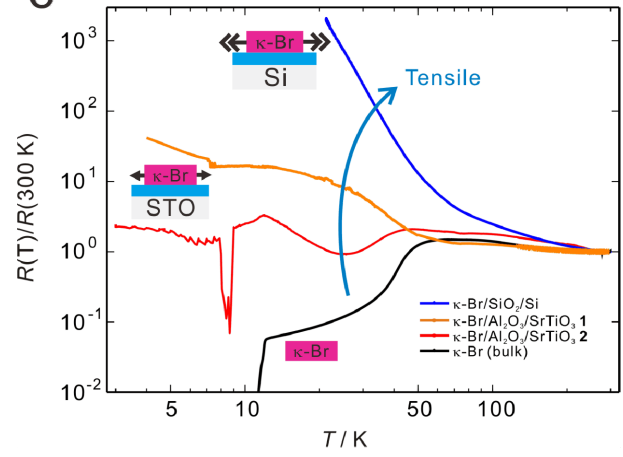

Fig. 1

Fig. 2

Fig. 3

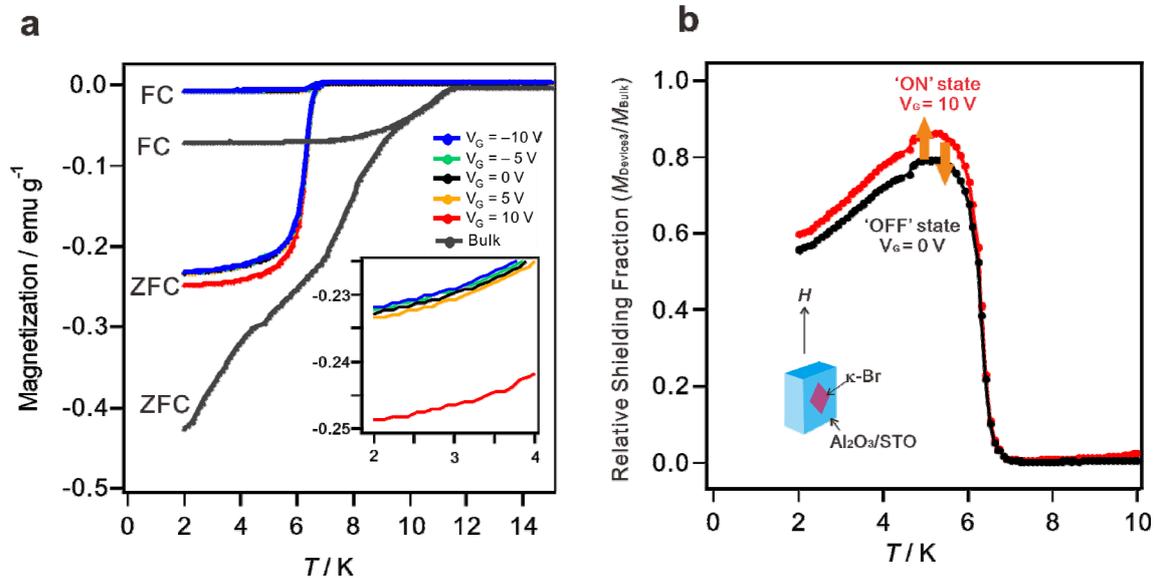

Fig. 4

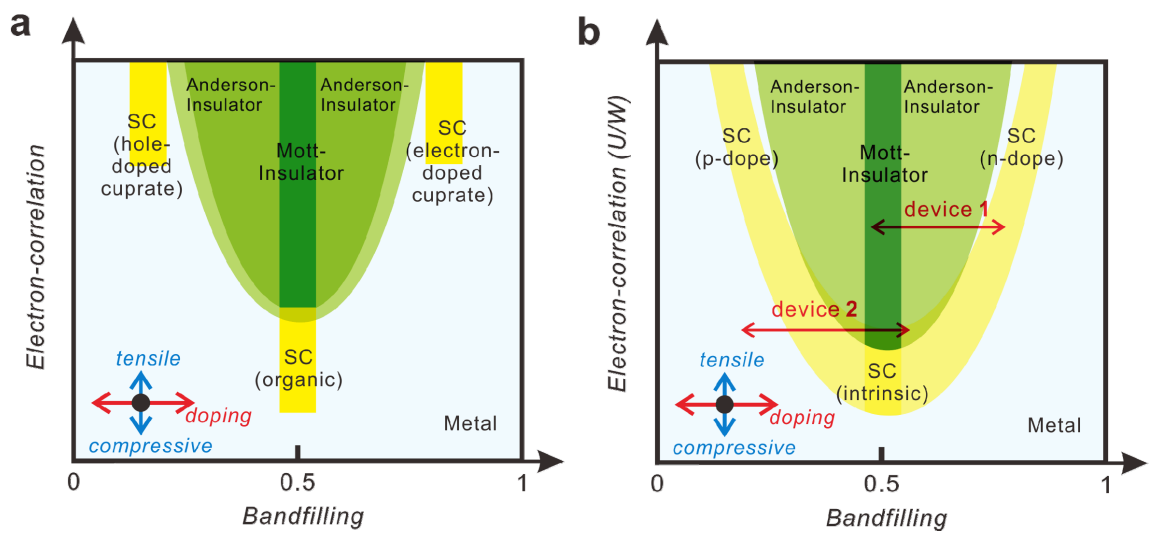

Fig. 5